# Kinetic Turbulence in the Terrestrial Magnetosheath: Cluster Observations

S. Y. Huang[1], F. Sahraoui[2], X. H. Deng[1,3], J. S. He[4], Z. G. Yuan[1], M. Zhou[3], Y. Pang[3], H. S. Fu[5]

[1]*School of Electronic and Information, Wuhan University, Wuhan, China.*

[2]*Laboratoire de Physique des Plasmas, CNRS-Ecole Polytechnique-UPMC, Palaiseau, France*

[3]*Institute of Space Science and Technology, Nanchang University, Nanchang, China*

[4]*School of Earth and Space Sciences, Peking University, Beijing, China.*

[5]*Space Science Institute, School of Astronautics, Beihang University, Beijing, China.*

***Abstract***

We present a first statistical study of subproton and electron scales turbulence in the terrestrial magnetosheath using the Cluster Search Coil Magnetometer (SCM) waveforms of the STAFF instrument measured in the frequency range [1,180] Hz. It is found that clear spectral breaks exist near the electron scale, which separate two power-law like frequency bands referred to as the dispersive and the electron dissipation ranges. The frequencies of the breaks $f_b$ are shown to be well correlated with the electron gyroscale $\rho_e$ rather than with the electron inertial length $d_e$. The distribution of the slopes below $f_b$ was found to be narrow and peaks near -2.9, while that of the slopes above $f_b$ was found broader, peaks near -5.2 and has values as low as

-7.5. This is the first time that such steep power-law spectra are reported in space plasma turbulence. These observations provide strong constraints on theoretical modeling of kinetic turbulence and dissipation in collisionless magnetized plasmas.

## *Introduction*

Turbulence is ubiquitous in astrophysical plasmas such as accretion disks, the interstellar medium, the near-Earth space [1-4], and laboratory plasmas such as those of fusion devices [5]. Turbulence plays a fundamental role in mass transport, energy dissipation and particle acceleration or heating, in particular at kinetic scales [6]. Therefore, it is crucial to determine the properties of kinetic turbulence (e.g. scaling law, anisotropy) experimentally, which should help to better understand the actual processes of energy dissipation (e.g., wave-particle interactions, coherent structures).

In the solar wind (SW), turbulence has been studied for decades in particular at MHD scales ($L \gg \rho_i \sim 100$km, the ion gyroradius) using data from different spacecraft (e.g., Voyager, Wind, Helios). In recent years the interest of the space community has shifted toward kinetic scales, i.e., subproton and electron scales, where key processes such as energy dissipation and particles heating occur [7-13]. Parallel to these observational work to determine the properties of turbulence at kinetic scales, large theoretical and numerical efforts have been done to tackle this challenging problem [6,14-19]. Despite all those efforts, several aspects of kinetic scale turbulence remain very controversial such as: i) the nature of the plasma modes that carry the turbulence cascade: Kinetic Alfvén Wave (KAW) [6,9,12-14] or whistler and/or other type of turbulence [16-20]; ii) the nature of the dissipative processes: resonant wave-particle interactions [13,14], coherent-like dissipation –e.g., magnetic reconnection [21-23]; iii) the actual scaling of the magnetic energy spectra, exponential [26] or power-law [12,13].

From the observational view point, the main obstacle that prevents one from fully addressing the previous controversies is the difficulty to measure the low amplitude electric and magnetic fluctuations in the SW at electron scales, because of the limited sensitivity of the current wave instruments [24,25]. Despite this limitation a few statistical studies have been conducted recently focusing on data intervals where the magnetic fluctuations have a high Signal-to-Noise-Ratio (SNR) [25,26]. In a ten years survey of the Cluster SCM burst mode waveforms in the SW, *Sahraoui et al.* [25] reported that spectral breaks occur near the Taylor-shifter electron gyroscale $f_{\rho e}$ followed by steeper power-law spectra. The distribution of the slopes below $f_{\rho e}$ was found narrow and peaked around -2.8, while that of the slopes above $f_{\rho e}$ was broader and peaked near -4. The break model used in *Alexandrova et al.* [26] yield similar results for 30% of the analyzed spectra, while the rest of the spectra were shown to be better fit with a curved (exponential) model (see discussion in [25]).

Here we propose to analyze another collisionless magnetized space plasma that is the terrestrial magnetosheath, i.e. part of the SW downstream of the bow shock. We take advantage of the high SNR available in the magnetosheath to overcome the limitation discussed above and to probe into the electron scales. In comparison with SW turbulence, magnetosheath turbulence is poorly known and only a handful of studies have been carried out in recent years [e.g. 3,27-34]. Two main similarities with SW turbulence emerged from those studies: i) A strong anisotropy ($k_{//}<<k_{\perp}$) both at subproton and electron scales [3,29,34], ii) The presence of kinetic instabilities and nonlinear structures [3,28,31]. Major differences exist though, e.g., i) magnetosheath turbulence evolves in a "confined" space limited by the bow shock and the magnetopause; these boundaries were shown to influence the anisotropy of the turbulence [3,32][a]; ii) the dominance of compressible fluctuations (e.g. mirror modes [3]).

In this letter, we use the waveforms data measured by the STAFF instrument onboard the Cluster spacecraft [35] to investigate the properties of terrestrial magnetosheath turbulence at electron scales. We selected 71 time intervals of 10 min during the period 2003-2007 when the SCM was in burst mode (BM). This allows us to study the frequency range [1, 180] Hz, which covers both subproton and electron scales.

***Observations***

Figure 1 shows the SNR ($SNR=10\cdot\log_{10}(dB^2/dB_{sens}^2)$) of the total magnetic energy spectra at 30 Hz for all the selected events (30Hz corresponds roughly to $k\rho_e\sim1$ for $V_f\sim250$ km/s, $B\sim14$ nT, $T_e\sim30$ eV using the Taylor hypothesis). One can see clearly that the SNRs of most of events (86%) are larger than 10, which contrasts with the low rate (16%) found in a survey of ten years data in the SW [25] (here, one event refers to a spectrum computed over 9.1s).

[a] It was shown in [3] that turbulence is non-axisymmetric around $\mathbf{B_0}$ when the normal to the magnetopause lies in the plane perpendicular to $\mathbf{B_0}$. A similar effect has been observed in SW turbulence [36].

Figure 2 shows an example of the studied observations measured by different Cluster instruments [37-39]: the electron spectrograms showing energy fluxes of ~ [0.1,10]keV, magnetic field and plasma density measurements showing relatively large fluctuations ($\delta B/<B>\sim 0.3$, $\delta n/<n>\sim 0.25$).

Figure 3 shows the histograms of the mean magnetosheath plasma parameters. Interestingly, these parameters show that the magnetosheath allows one to cover a broad range of physical conditions in a collisionless magnetized plasma: electron plasma $\beta_e \in [0.02, 21]$), ion cyclotron frequency $f_{ci} \in [0.04, 1.0]$ Hz, plasma flow $V_{flow} \in [70, 480]$ km/s; ratio of ion to electron temperatures $T_i/T_e \in [0.3, 24]$. Were excluded from the original list of events (Figure 1) all spectra found to have bumps (or "knees") at high frequency and those that did not show clear breaks (this explains the small number of events in Figure 1 compared to Figure 3). The "knees" might be caused by quasi-parallel whistler or beam driven modes [30].

In Figure 4 are presented some examples of the analyzed spectra. They show clear spectral breaks near the Taylor-shifted electron gyroradius ($f_{\rho e}=V_{flow}/2\pi\rho_e$) or inertial length ($f_{de}=V_{flow}/2\pi d_e$), which separate two well defined frequency bands having power-law like scaling. To test which scale, $\rho_e$ or $d_e$, that would correspond statistically the observed breaks, we plotted in Figure 5a-5b the correlations of the break frequencies with $f_{de}$ and $f_{\rho e}$. A higher correlation is found between the break frequencies and $\rho_e$ (0.73) rather than $d_e$ (0.23). This result is similar to the observations of SW turbulence [12,13,25], although here we found a lower correlation with $d_e$ than in SW observations [25]. This may be explained by the large number of events found here with β>1 in the present study compared to the SW case. The strong correlation of the spectral breaks with the electron gyroscale suggests that this latter plays the role of a dissipation scale in magnetosheath turbulence, similarly to SW turbulence [12,25].

To investigate the scaling of magnetosheath turbulence, we fit the spectra with a double-power-law model below and above the spectral breaks [25]. The histograms of the slopes are shown in Figure 5c-5d. The slopes below the spectral breaks cover the range ~[-3.5, -2.4] with a peak near -2.9. These values are consistent with previous magnetosheath observations [31,33] and with SW observations [7,8,12,13,25].The spectra above the spectral breaks are, however, steeper than those of the SW; they have slopes distributed in the range ~[-4, -7.5]. To our knowledge these are the steepest spectra ever reported in space plasmas. The bulk of the distribution (peaked near -5.2) is nevertheless in agreement with several existing theoretical predictions [6,15,16,19].

To investigate the reason as to why such steep spectra have not been reported in SW observations we test the possible role of the limited SNR in the SW compared to the case of the magnetosheath. In Figure 6 we plotted the correlation between the slopes of the spectra above the electron spectral break and the SNR. As one can see, a similar correlation coefficient (C~0.47) is found as in the SW (C=0.53 [25]) for all events that have SNR>10. However, as we increase the threshold on the SNR we observe a decreasing correlation: C~0.19 and C~0.1 for SNR>20 and SNR>25, respectively. We can see furthermore that slopes<-6 correspond to events with SNR>20, which were extremely rare in the SW [25]. From the above observations one may conclude that SNR higher than 20 are needed in the SW to fully address the scaling on the magnetic energy spectra at electron scales. This result sets up an instrumental requirement that the future space mission dedicated to SW turbulence need to fulfill. We note that we also investigated the possible dependence of the scaling on the plasma parameters, such as ion $\beta_e$, and found no significant low correlation (not shown here).

Above we used the Taylor assumption to link the observed breaks in frequency with the spatial scales of ions and electrons, similarly to what has been used in previous magnetosheath [30] and SW studies [12,25,6]. While this assumption is generally valid at MHD scales (where $M_A=V_{sw}/V_A>>1$, the Alfvén Mach number) it may fail at electron scales (both in the W of magnetosheath) if high frequency or dispersive modes are present (e.g. whistlers). Sahraoui et al. [17] have proposed a simple test based on estimating the ratio between the break frequencies of electron to ion (formula 15 therein). This test was performed on the present data (not shown here) and showed that most of the intervals used here reasonably fulfill the Taylor assumption even at electron scales. A complete study of this problem is being published elsewhere [Huang et al., 2014].

***Conclusions***

Strong controversies exist about the scaling and the nature of the turbulence at electron scales in the SW. Due the low amplitude fluctuations of the electric and the magnetic fields in the SW, the current sensitivity of the instruments do not allow one to fully address electron scale turbulence, and thus to remove part of the existing controversies. Magnetosheath turbulence presents another alternative to explore electron scale turbulence in collisionless magnetized plasmas. The statistical results shown above confirm the presence of the spectral breaks near the electron gyroscale $\rho_e$, followed by steep power-law like spectra with slopes as high as -7.5. This suggests that $\rho_e$ plays the role of the dissipation scale, as previously found in SW turbulence. The steepest spectra reported here have not been predicted so far by any existing theoretical or numerical studies. The present results provide strong observational constraints on the current theoretical efforts to understand the problem of energy cascade and dissipation in collisionless magnetized plasmas. Some important questions, such as the nature of the plasma modes involved in the cascade at electron scales and the degree of anisotropy of the turbulence, will be investigated in future studies.

***Acknowledge***

The data used in this work come from the ESA/Cluster Active Archive (CAA) and AMDA (IRAP, France). This research was supported by the National Natural Science Foundation of China (NSFC) under grants 40890163, 41174147 and 41004060. F. Sahraoui acknowledges support from the THESOW project funded by ANR (Agence

Nationale de la Recherche).

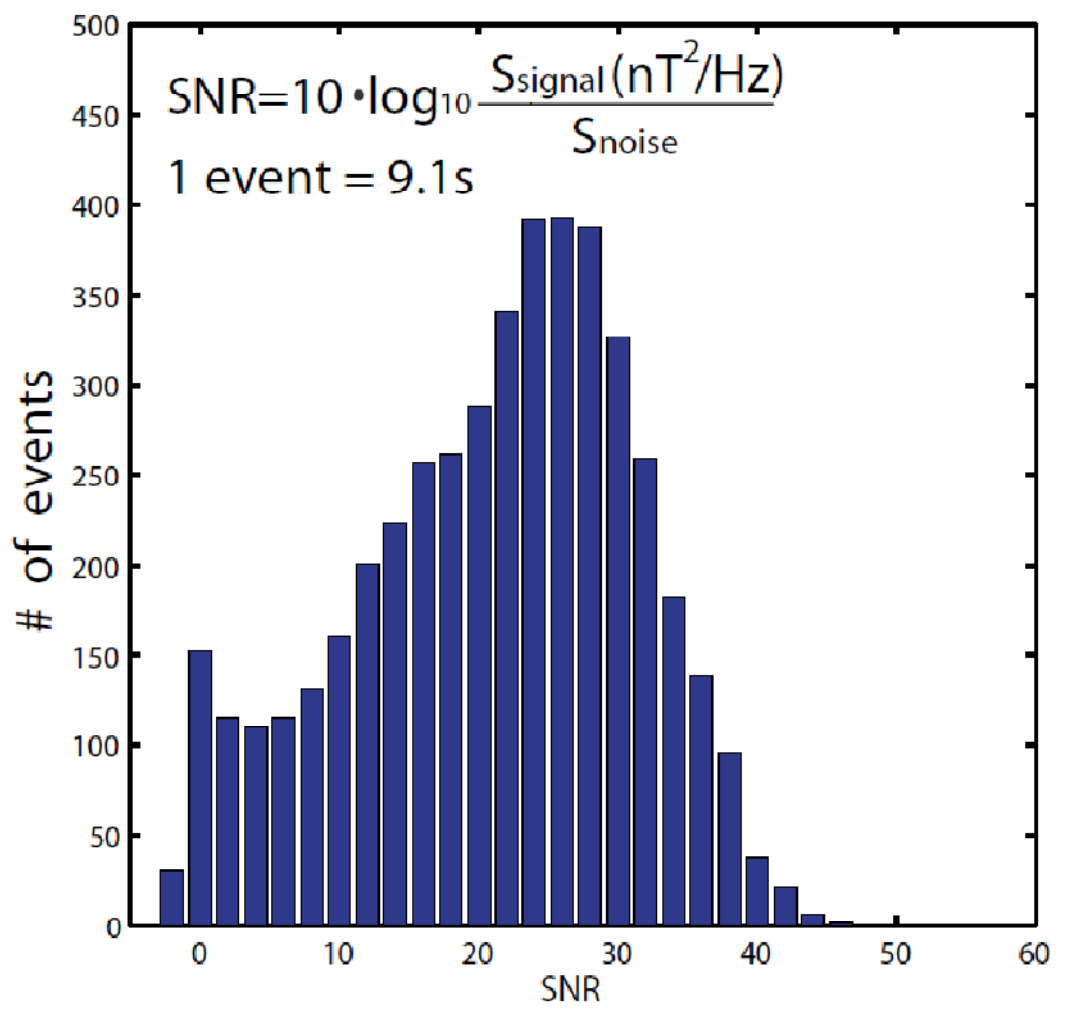

**Fig.1.** SNR of total magnetic energy spectra at 30 Hz in the terrestrial magnetosheath turbulence.

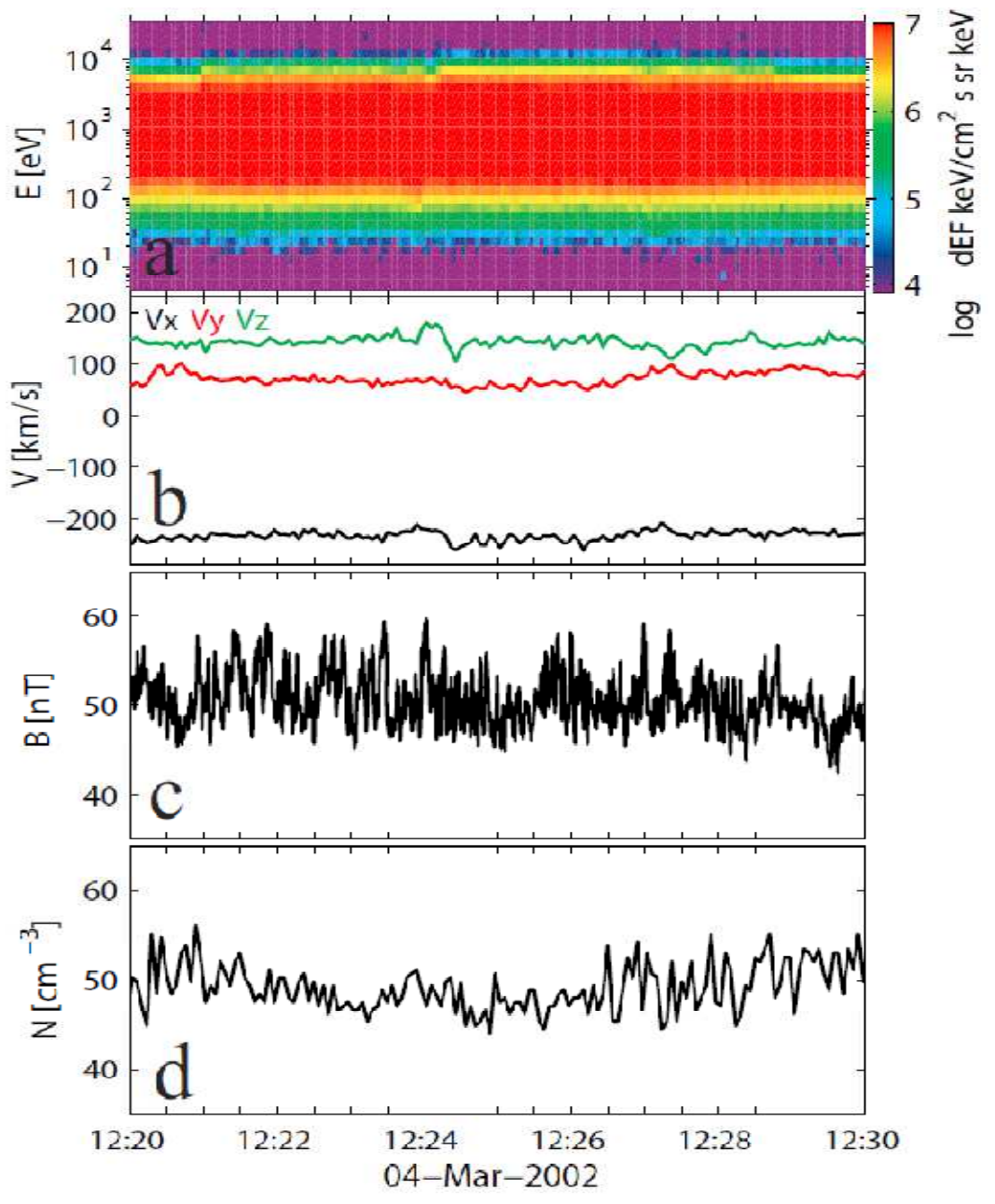

**Fig. 2.** Typical example of the studied Magnetosheath data: (a) electron spectrogram; (b) plasma flow velocity; (c) magnitude of the magnetic field; (d) ion density

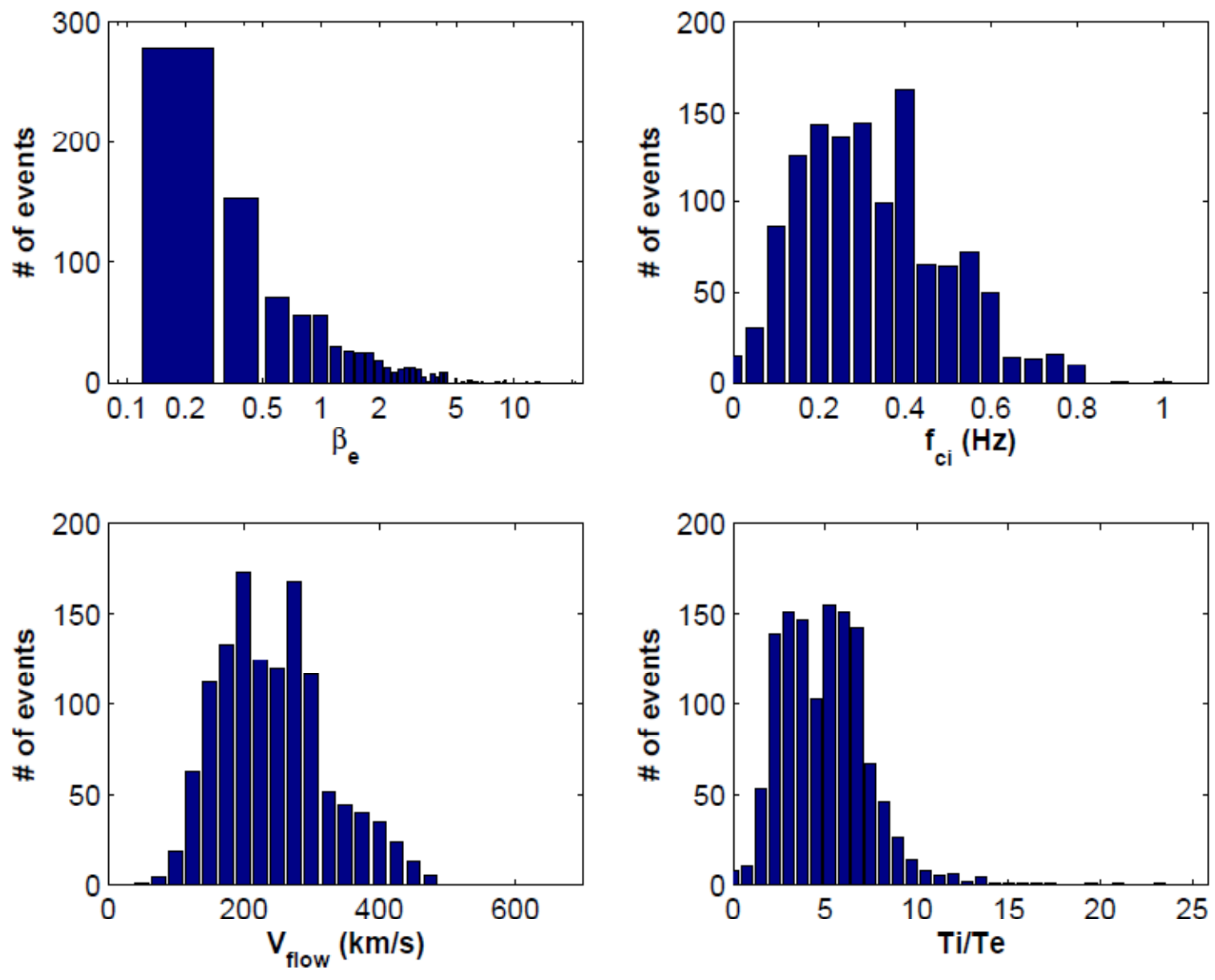

**Fig. 3.** The histograms of the mean plasma parameters for the magnetosheath observations: (a) electron plasma βe; (b) ion gyrofrequency; (c) plasma flow speed; (d) the ratio of ion and electron temperatures.

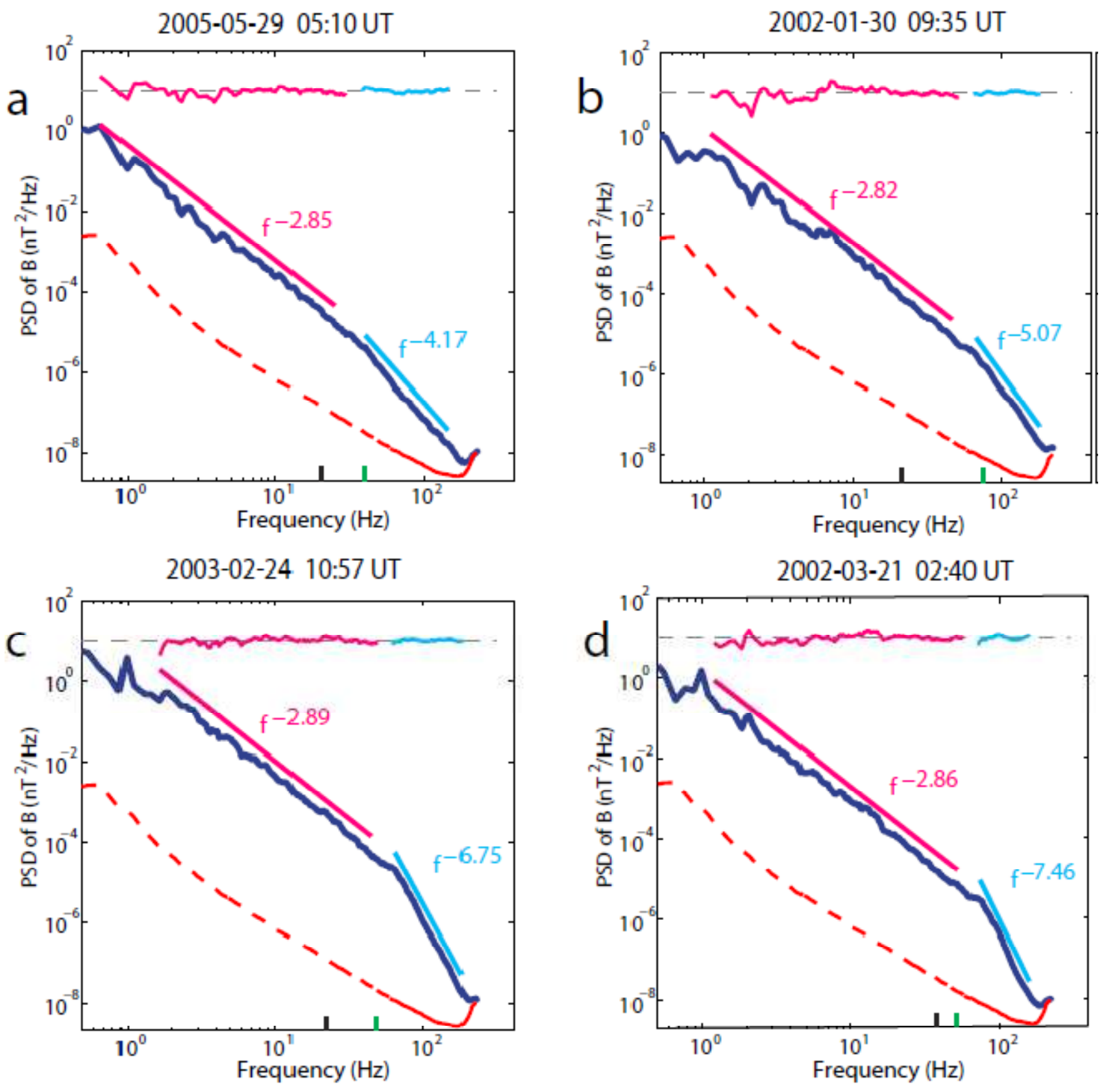

**Fig.4.** Examples of analyzed magnetic spectra. The red dashed curve is the in-flight sensitivity of STAFF SCM instrument. The vertical black and green lines are the Taylor-shifted frequencies of electron gyroradius scale and electron inertial length. The horizontal red and cyan lines are the compensated spectra showing the quality of the power-law fits.

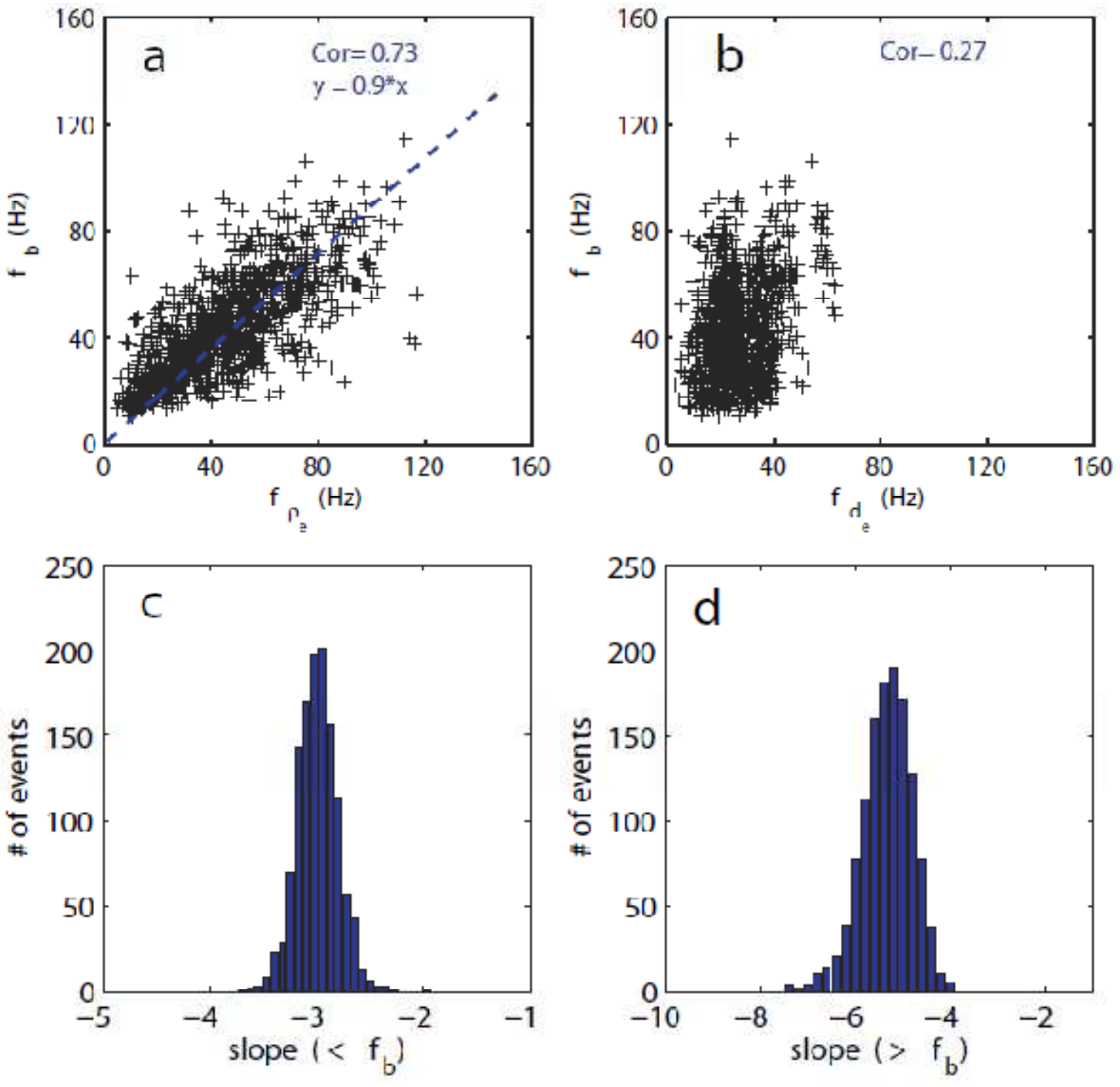

**Fig. 5.** Correlations of the frequencies of the spectral breaks with (a) the Taylor-shifted electron gyroradius and (b) electron inertial length scale. Linear fits with corresponding functions are shown by blue dashed lines. Histograms of the spectral indices above (c) and below (d) the electron spectral breaks.

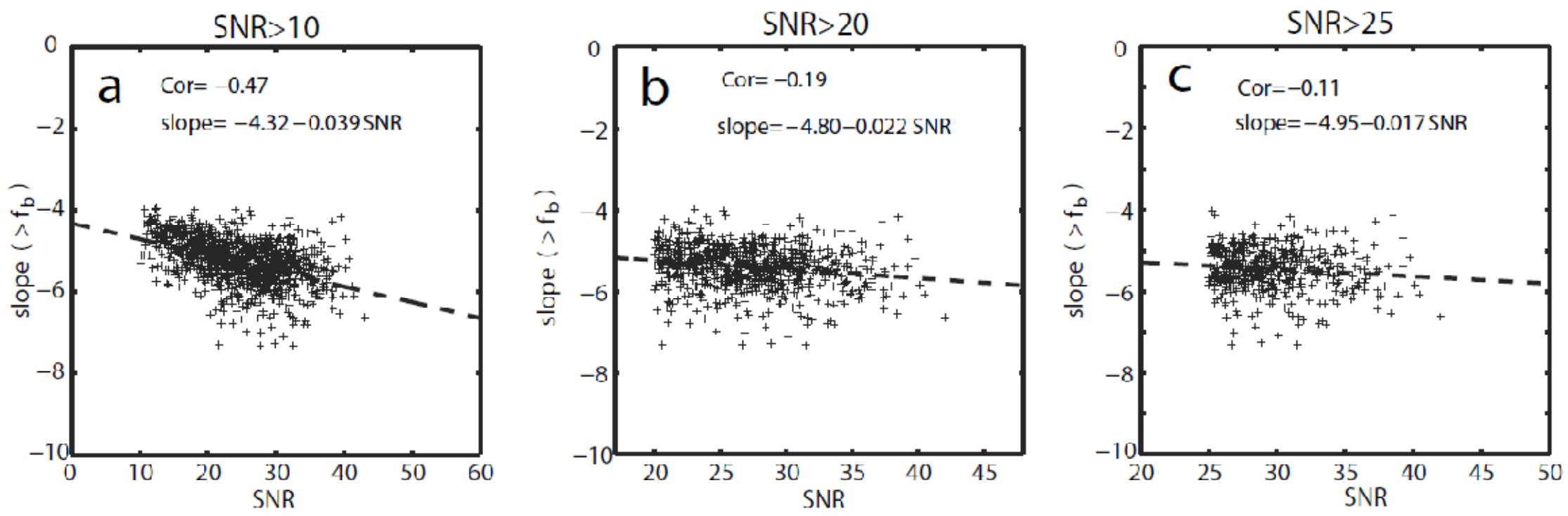

**Fig. 6.** Correlation between the SNR and the slopes of the spectra above the spectral break $f_{be}$ for the events having SNR>10 (a), SNR>20 (b) and SNR>25 (c).